\documentclass{article}
\usepackage{hyperref}
\usepackage{amssymb}
\newcommand{\R}{\mathbb{R}}
\newcommand{\eps}{\varepsilon}
\title{Lower bound on the Voronoi diagram of lines in~\(\R^d\)}
\author{Marc Glisse\footnote{Université Paris-Saclay, CNRS, Inria, Laboratoire de Mathématiques d'Orsay, 91405, Orsay, France}}
\begin{document}
\maketitle
\abstract{This note gives a lower bound of $\Omega(n^{\lceil 2d/3\rceil})$ on the maximal complexity of the Euclidean Voronoi diagram of $n$ non-intersecting lines in $\R^d$ for $d>2$.}

\section{Introduction}
One of the most common objects in computational geometry is the Voronoi diagram of a set of sites, which partitions the space into regions according to which site is the closest. The most basic case is when the sites are $n$ points in $\R^d$, where the maximal complexity of the diagram is known to be $\Theta(n^{\lceil d/2\rceil})$. The next simplest object after a point is a line, which brings us to one of the most famous open problems in computational geometry: what is the maximal complexity of the Voronoi diagram of $n$ lines in $\R^3$? This is only known to be somewhere between $\Omega(n^2)$ and $O(n^{3+\varepsilon})$ (valid for any $\varepsilon>0$), quite a large gap. The goal of this paper is not to attack this very hard problem, but to consider what happens for lines in $\R^d$ with $d>3$. This question is much less studied, but has already been mentioned a few times~\cite{ARONOV2002183,stateunion}. Looking at the Voronoi diagram as a lower envelope in $\R^{d+1}$ yields an upper bound of $O(n^{d+\varepsilon})$. The only lower-bounds we are aware of are $\Omega(n^2)$ (the construction in $\R^3$ trivially embeds in $\R^d$ for $d\geq3$) and $\Omega(n^{\lfloor d/2\rfloor})$ (place points in $\R^{d-1}$ that maximize the size of the Voronoi diagram, then extend them all into parallel lines in $\R^d$ along the extra dimension). In this note we give a slightly less pessimistic lower bound.

\section{Basic construction}
An important property of the lines is whether they are allowed to intersect or not. In $\R^2$, intersecting lines may form a square grid, and the Voronoi diagram has $\Theta(n^2)$ cells. On the other hand, non-intersecting lines must be parallel and their Voronoi diagram has linear complexity. Our approach will be first to allow intersections, and then notice that in dimension 3 or higher, the lines can be perturbed infinitesimally without reducing the complexity of the diagram. Our starting point is thus that in $\R^2$ the Voronoi diagram of lines can have quadratic complexity.

One possible intuition is that a grid of $n$ lines in a plane is somewhat similar to a grid of $n^2$ points, and we can then think where to place plane patches to get a large Voronoi diagram of point sites, which has been considered~\cite{polyhedron} before.

\subsection{Aggregation}
Assume we have a set $A$ of $n$ lines in $\R^p$ whose Voronoi diagram has complexity $C_1$\footnote{As an abuse of language, we may call complexity the number of tuples of sites that define at least one Voronoi cell. This is a lower bound on the usual complexity and is thus fine for our purpose.}, and a set $B$ of $n$ lines in $\R^q$ whose Voronoi diagram has complexity $C_2$. We can trivially embed the points from the first space into $\R^{p+q+1}$ as $(x_1,\ldots,x_p,0,\ldots,0)$ and the points from the second space as $(0,\ldots,0,y_1,\ldots,y_q,1)$. We now argue that the resulting set of $2n$ lines has a Voronoi diagram of complexity at least $C_1C_2$.

A point $(x_1,\ldots,x_p)$ is equidistant to some subset $A'$ of $A$ in $\R^p$ and further from the other lines of $A$, depending on the Voronoi cell it belongs to. This remains true for any point in $\R^{p+q+1}$ whose first $p$ coordinates are $(x_1,\ldots,x_p)$. Similarly, a point $(y_1,\ldots,y_q)\in\R^q$ is equidistant from a subset $B'$ of $B$ and further from the other lines of $B$, and this remains true in $\R^{p+q+1}$ for any point whose coordinates $p+1$ to $p+q$ are $(y_1,\ldots,y_q)$. The points $P_t$ of coordinates $(x_1,\ldots,x_p,y_1,\ldots,y_q,t)$ for $t\in\R$ form a line. When $t$ tends to $-\infty$, $P_t$ is closer to any line of $A$ than to any line of $B$. Symmetrically, when $t$ tends to $+\infty$, $P_t$ is closer to any line of $B$ than to any line of $A$. By continuity, there exists a point $P_{t^*}$ that is equidistant to $A'$ and $B'$ (and further from the other lines). The tuple $A'\cup B'$ thus defines (at least) one Voronoi cell, and there are at least $C_1C_2$ Voronoi cells.

Starting from a complexity $n^2$ when $d=2$, this builds a Voronoi diagram of complexity $n^{2c}$ in dimension $3c-1$, and in particular $n^4$ in $\R^5$.

\subsection{Perturbation}
The fact that we can perturb the lines without destroying the cells defined above should be intuitive, but we confirm it with an explicit scheme in $\R^5$. $\eps$ represents a positive number smaller than $\frac1{8n}$.

We define four sets of lines in $\R^5$ indexed by integers $i$, $j$, $k$, $l$ between $1$ and $n$.
Line $A_i$ has coordinates $(i,*,0,0,0)$ ($*$ represents a free coordinate),
line $B_j$ has coordinates $(*,j,\eps,0,0)$,
line $C_k$ has coordinates $(0,0,k,*,1)$,
line $D_l$ has coordinates $(\eps,0,*,l,1)$.
Thanks to $\eps$, these lines do not intersect.

Now we consider points $x$ on the line $E_{i,j,k,l}$ whose coordinates are
$x_1=i$,
$x_2=j+\sqrt{2k\eps-\eps^2}$,
$x_3=k$,
$x_4=l+\sqrt{2i\eps-\eps^2}$,
$x_5=*$.

They are equidistant from $A_i$ and $B_j$ since $(x_1-i)^2+x_3^2=(x_2-j)^2+(x_3-\eps)^2$ and all the other lines of $A$ and $B$ are further.

Similarly, they are equidistant from $C_k$ and $D_l$ and the other lines of $C$ and $D$ are further.

Walking on $E_{i,j,k,l}$, for $x_5=-\infty$, $x$ is closer to $A$ and $B$ than to $C$ and $D$, and vice versa for $x_5=+\infty$, so there is an intermediate $x_5$ such that all 4 lines are equidistant (we can even compute it explicitly as $x_5=\frac12(1+x_1^2+x_2^2-x_3^2-x_4^2)$). Since we have a Voronoi cell for each 4-tuple $(i,j,k,l)$, the Voronoi diagram has complexity $\Omega(n^4)$.

\subsection{Dimension $4$}
At this point, we have a quadratic bound in dimension $2$, which still applies in dimensions $3$ and $4$, and a bound of $n^4$ in dimension $5$. This jump is not very satisfactory. While we are not going to touch the 3d case here, a cubic bound in $\R^4$ is doable. We can define non-intersecting lines $A_i$: $(i,*,\eps,0)$, $B_j$: $(*,j,0,0)$ and $C_k$: $(*,0,k,1)$ for $i$, $j$ and $k$ between $1$ and $n$. Then the point of coordinates $(i+\sqrt{2\eps k-\eps^2},j,k,\frac{j^2-k^2+1}2)$ is equidistant from $A_i$, $B_j$ and $C_k$ and further from the other lines. This gives a lower bound of $n^3$. It can be aggregated with some grids as above to handle dimensions $3c+1$.

\section{Generalizations}
\subsection{$k$-flats}
After lines, a natural question is what happens for planes, or more generally $k$-flats. We can similarly build a grid of complexity $n^{k+1}$ in $\R^{k+1}$ (that intersects) and aggregate those grids to get a complexity $n^{c(k+1)}$ in $\R^{c(k+2)-1}$ (without intersection for $c>1$). The exponent of the complexity in $\R^d$ is of order $\frac{k+1}{k+2}d$, which interpolates between the known bounds for points and $(d-2)$-flats.

\subsection{Metric}
This construction does not strongly depend on the Euclidean metric, in particular it still works for $L^p$ if $1<p<\infty$. For $p=1$, the argument about the last coordinate tending to infinity does not work anymore, but can be fixed by replacing $1$ with something larger than $dn$ as last coordinate. For $p=\infty$, the construction may be too degenerate, but we expect similar constructions would yield the same bound.

\section{Conclusion}

We have seen that for $d>2$, the Voronoi diagram of $n$ non-intersecting lines in $\R^d$ can have complexity $\Omega(n^{\left\lceil\frac{2d}3\right\rceil})$, which is still far from the upper bound of $O(n^{d+\varepsilon})$. It is always tempting to conjecture that the new lower bound is closer to the right answer than the upper bound, but we do not wish to venture that far at this point.

For points, the lower bound is usually achieved by placing the points on the moment curve. It would be interesting to try and define some ruled surface or other parameterized smooth family of lines that achieves the same bound as presented here.

\section{Acknowledgement}
The author wishes to thank Nina Amenta for old, joint work on a related problem, and Boris Aronov for a short discussion on this topic.
\bibliographystyle{plainurl}
\bibliography{biblio}
\end{document}